\RequirePackage{fix-cm}
\documentclass[twocolumn,showpacs,preprintnumbers,amsmath,amssymb,showkeys,superscriptaddress]{revtex4}

\usepackage{graphicx}
\usepackage{textcomp}
\usepackage{nicefrac}
\usepackage{color}
\usepackage{amsmath}

\newcommand{\grfsize}{\fontsize{15pt}{15pt}\selectfont} 
\makeatletter

  \@ifundefined{ifGPcolor}{%
    \newif\ifGPcolor
    \GPcolorfalse
  }{}%
  \@ifundefined{ifGPblacktext}{%
    \newif\ifGPblacktext
    \GPblacktexttrue
  }{}%
  \let\gplgaddtomacro\g@addto@macro
  \gdef\gplbacktext{}%
  \gdef\gplfronttext{}%
  \makeatother
\begin{document}

\preprint{APS/123-QED}

\title{Measuring the deformation of a ferrogel sphere in a homogeneous magnetic field}%
\author{Christian Gollwitzer}
\affiliation{%
Experimentalphysik V, Universit\"at Bayreuth, 95440 Bayreuth, Germany
}%

\author{Alexander Turanov}%
\affiliation{%
Experimentalphysik V, Universit\"at Bayreuth, 95440 Bayreuth, Germany
}%

\author{Marina Krekhova}
\author{G\"unter Lattermann}
\affiliation{%
Makromolekulare Chemie I, Universit\"at Bayreuth, 95440 Bayreuth, Germany
}%
\author{Ingo Rehberg}
\author{Reinhard Richter}
\affiliation{%
Experimentalphysik V, Universit\"at Bayreuth, 95440 Bayreuth, Germany
}%

\date{\today}

\begin{abstract}
A sphere of a ferrogel is exposed to a homogeneous magnetic field. In
accordance to theoretical predictions, it gets elongated along the field lines.
The time-dependence of the elastic shear modulus causes the elongation to
increase with time analogously to mechanic creep experiments, and the rapid
excitation causes the sphere to vibrate. Both phenomena can be well described by
a damped harmonic oscillator model. By comparing the elongation along the field with the
contraction perpendicular to it, we can calculate Poisson's ratio of the gel. The magnitude of the
elongation is compared with the theoretical predictions for elastic spheres in homogeneous fields.
\end{abstract}

\maketitle
\section{Introduction}
A sphere is the most perfect, most symmetric geometrical object. The
Pythagoreans believed that the sun and the earth are perfect and thus
spherical. Of course the rotation of the Earth is breaking the spherical
symmetry. Whether this changes the shape of the Earth to a prolate or oblate
rotational ellipsoid has been the subject of a long lasting quarrel between
Newton and Cassini \cite{chandrasekhar1987}. The breaking of the spherical
symmetry is found to be important in various fields of physics. The efficiency
of nuclear fission depends on the shape of the nucleus \cite{bohr1939}. Particularly convenient is a breaking of the
symmetry by an external electric or magnetic field. A drop of a magnetic
liquid, a colloidal dispersion of magnetic nanoparticles \cite{rosensweig1985},
elongates along the direction of the applied homogeneous magnetic field
\cite{bacri1982,bacri1983,flament1996}. Very recently an amount of ``quantum
ferrofluid'', consisting of dipolar Cr-Atoms, has been found as well to elongate
in a magnetic field \cite{lahaye2007}.

All examples presented above have in common that they deal with fluid bodies,
without any elasticity. In contrast smart nanocomposite polymer gels
\cite{varga2003} are kept together by their elastic polymer matrix. Especially
ferrogels \cite{zrinyi1996a,zrinyi2000} are a promising class for many
applications, like soft actuators, magnetic valves, magnetoelastic mobile
robots \cite{zimmermann2006,zimmermann2007}, artificial muscles
\cite{babincova2001}, or magnetic controlled drug delivery~\cite{lao2004}. All these
applications are based on the magnetic deformation effect, a coupling
between the mechanic and magnetic degrees of freedom.

The magnetic deformation effect is studied in its simplest form for a spherical gel body subjected to a
homogeneous magnetic field. An approximation for the resulting deformation has
been given in 1960 by Landau for the case of a dielectric elastic sphere
\cite{landau1960_8} and can readily be transferred to ferrogels
\cite{raikher2003,raikher2005,raikher2005b}.
The relative elongation $\varepsilon$ in this case was calculated as 
\begin{equation}
\varepsilon = \frac{\kappa\mu_0}{G} M^2 \label{eq:relelong}
\end{equation}
where $\kappa=\nicefrac{1}{15}$, $G$ is the shear modulus and $M$ is the magnetization.
Recently, the elongation has been recomputed by
Raikher \cite{raikher2005} without constraining the shape to an ellipsoid. In
this case, the elongation is expected to be $30\,\%$ larger. This effect has not yet been
observed, a possible reason being that it is rather small for the large values
of $G$, characteristic for most of the covalently cross linked polymer gels
~\cite{zrinyi1996a,zrinyi1996b}. In contrast, the elasticity of the new class of
thermoreversible ferrogels ~\cite{lattermann2006} can reversibly be tuned via
their temperature.
In the present paper, we cast thermoreversible ferrogels in spherical
samples and expose them to a uniform magnetic field, to test the above predictions.

\section{Experiment}
\begin{figure}
\includegraphics[width=0.6\linewidth]{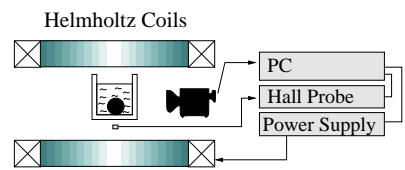}
\caption{Scheme of the experimental setup.}
\label{fig:setup}
\end{figure}

\subsection{Setup}

Our experimental setup is sketched in Fig.\,\ref{fig:setup}. A ferrogel ball is immersed in a
rectangular container, positioned on the common axis midway between two Helmholtz coils. For the
empty Helmholtz pair of coils, the spatial homogeneity is better than $\pm 1\,\%$. This grade is
valid within a cylinder of $1\,\mathrm{cm}$ in diameter and $14\,\mathrm{cm}$ in height oriented
symmetrically around the center of the coils. The coils are powered by a current amplifier (fug
electronic GmbH), which is controlled by a computer. The magnetic system cannot follow the control signal immediately. For a maximal
jump height $\Delta B=36\,\mathrm{mT}$ the field is reached after $t_B=30\,\mathrm{ms}$, as
recorded by the Hall probe (Group3-LPT-231) connected to a digital teslameter (DTM 141). 

\begin{figure}
\includegraphics[width=0.5\linewidth]{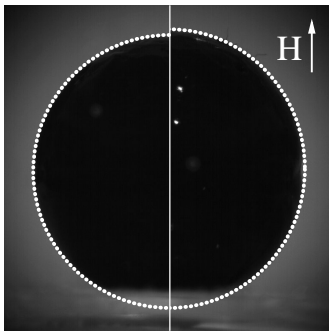}
\caption{Image of the symmetrical~(left) and distorted ball~(right). 
The dotted line displays a fit with a circle~(left) and an ellipse~(right).}
\label{fig:ball}
\end{figure}

The temporal evolution of the ball shape is recorded using a highspeed
camera, capable of taking $400$ frames per second with a resolution of $768\times768$ pixels.
Figure~\ref{fig:ball} shows the original and distorted shape. The dotted curves stem from a fit of 
a circle to the edges in the image. The circle is located utilizing a normalized correlation
technique~\cite{dickey1991}, where the correlation between the gradient of the
image and a circle is maximized. The gradient is estimated by the Sobel operator~\cite{song2006}, and the circle is
rasterized with anti-aliasing provided by a Gaussian filter~\cite{crow1977}. This method is able to
extract the radius and the coordinates of the center with subpixel resolution. 

\subsection{Material}
The ball was prepared with a magnetic, thermoplastic elastomer gel or in other words with a
thermoreversible ferrogel~\cite{lattermann2006}. Therefore, an ABA-type
poly(styrene-b-(ethylene-co-butylene)-b-styrene) (SEBS) triblock copolymer (Kraton~G-1650) was used
as a gelator. As measured with size exclusion chro\-ma\-to\-gra\-phy (SEC), Kraton~G-1650 exhibits a
molar mass of  $99\,000$. The styrene content is $29\,\%$ (manufacturer information). The gelator
concentration is $3.5\,\mathrm{w}\%$ per paraffin oil used. The ferrofluid was prepared in the
classical way~\cite{lattermann2006} with $24.5\,\mathrm{w}\%$ magnetite per paraffin oil Finavestan
A~50\,B (Total). The shear viscosity at $20\,$\textdegree C of $11\,\mathrm{mPa\,s}$ and the molar
mass of $280$ (manufacturer information) are lower than that of the earlier used paraffin oil
Finavestan A~80\,B~\cite{lattermann2006}. Using A~80\,B, the resulting ferrogels are too stiff to be
remarkably deformed by the magnetic field. On the other hand, using the gelator Kraton G-1652 with a
lower molar mass of $79\,000$ (SEC), the prepared ferrogels are still softer, i.\,e.\ they are even
less shape-retaining. Furthermore, they sweat out ferrofluid, slowly on standing, faster in the
magnetic field.

\subsection{Sample preparation}
Above its softening temperature at
$45$\textdegree{}C the material becomes a magnetic liquid.
We produce the ferrogel sphere by casting the liquid into an aluminum mould at 
$55$\textdegree{}C. The
mould consists of two parts with a spherical cavity, tightly screwed together,
which is connected by a thin channel to a hopper mounted on top of the upper
part. To avoid any air bubbles in the final sphere, the mould is first filled
with the liquefied material under vacuum ($\approx 1\,\mathrm{mPa}$). Then atmospheric pressure is
applied, which compresses any low-pressure air bubbles. By repeating the process of
varying the ambient pressure and under the influence of gravity, eventually all
bubbles leave the mould via the hopper.
Next we cool down the mould and separate the two parts of it.

The ball is then
immersed into water with a temperature of 24\textdegree{}C, where all experiments have been carried out.
The water contains an amount of salt, so that the density of the liquid is only
slightly less than that of the ball. This is necessary to reduce the influence
of the gravity on the shape of the sample; the ball is so soft that, without buoyancy,
it is deformed into an oblate ellipsoid under gravity. As an additional benefit we can easily
measure the density of the gel by examining the density of the fluid.
The density determined by this method is $\rho = 1.085\pm0.005\,\mathrm{g/cm^3}$.

\begin{figure}
\resizebox{0.8\columnwidth}{!}{\grfsize
\begingroup 
  \gdef\gplbacktext{}%
  \gdef\gplfronttext{}%

    \ifGPblacktext
    \def\colorrgb#1{}%
    \def\colorgray#1{}%
  \else
    \ifGPcolor
      \def\colorrgb#1{\color[rgb]{#1}}%
      \def\colorgray#1{\color[gray]{#1}}%
      \expandafter\def\csname LTw\endcsname{\color{white}}%
      \expandafter\def\csname LTb\endcsname{\color{black}}%
      \expandafter\def\csname LTa\endcsname{\color{black}}%
      \expandafter\def\csname LT0\endcsname{\color[rgb]{1,0,0}}%
      \expandafter\def\csname LT1\endcsname{\color[rgb]{0,1,0}}%
      \expandafter\def\csname LT2\endcsname{\color[rgb]{0,0,1}}%
      \expandafter\def\csname LT3\endcsname{\color[rgb]{1,0,1}}%
      \expandafter\def\csname LT4\endcsname{\color[rgb]{0,1,1}}%
      \expandafter\def\csname LT5\endcsname{\color[rgb]{1,1,0}}%
      \expandafter\def\csname LT6\endcsname{\color[rgb]{0,0,0}}%
      \expandafter\def\csname LT7\endcsname{\color[rgb]{1,0.3,0}}%
      \expandafter\def\csname LT8\endcsname{\color[rgb]{0.5,0.5,0.5}}%
    \else
      \def\colorrgb#1{\color{black}}%
      \def\colorgray#1{\color[gray]{#1}}%
      \expandafter\def\csname LTw\endcsname{\color{white}}%
      \expandafter\def\csname LTb\endcsname{\color{black}}%
      \expandafter\def\csname LTa\endcsname{\color{black}}%
      \expandafter\def\csname LT0\endcsname{\color{black}}%
      \expandafter\def\csname LT1\endcsname{\color{black}}%
      \expandafter\def\csname LT2\endcsname{\color{black}}%
      \expandafter\def\csname LT3\endcsname{\color{black}}%
      \expandafter\def\csname LT4\endcsname{\color{black}}%
      \expandafter\def\csname LT5\endcsname{\color{black}}%
      \expandafter\def\csname LT6\endcsname{\color{black}}%
      \expandafter\def\csname LT7\endcsname{\color{black}}%
      \expandafter\def\csname LT8\endcsname{\color{black}}%
    \fi
  \fi
  \setlength{\unitlength}{0.0500bp}%
  \begin{picture}(7200.00,5040.00)%
    \gplgaddtomacro\gplbacktext{%
      \csname LTb\endcsname%
      \put(1092,840){\makebox(0,0)[r]{\strut{} 0}}%
      \put(1092,1226){\makebox(0,0)[r]{\strut{} 1}}%
      \put(1092,1613){\makebox(0,0)[r]{\strut{} 2}}%
      \put(1092,1999){\makebox(0,0)[r]{\strut{} 3}}%
      \put(1092,2386){\makebox(0,0)[r]{\strut{} 4}}%
      \put(1092,2772){\makebox(0,0)[r]{\strut{} 5}}%
      \put(1092,3158){\makebox(0,0)[r]{\strut{} 6}}%
      \put(1092,3545){\makebox(0,0)[r]{\strut{} 7}}%
      \put(1092,3931){\makebox(0,0)[r]{\strut{} 8}}%
      \put(1092,4318){\makebox(0,0)[r]{\strut{} 9}}%
      \put(1092,4704){\makebox(0,0)[r]{\strut{} 10}}%
      \put(1260,560){\makebox(0,0){\strut{} 0}}%
      \put(2353,560){\makebox(0,0){\strut{} 5}}%
      \put(3446,560){\makebox(0,0){\strut{} 10}}%
      \put(4538,560){\makebox(0,0){\strut{} 15}}%
      \put(5631,560){\makebox(0,0){\strut{} 20}}%
      \put(6724,560){\makebox(0,0){\strut{} 25}}%
      \put(280,2772){\rotatebox{90}{\makebox(0,0){\strut{}Magnetization M (kA/m)}}}%
      \put(3992,140){\makebox(0,0){\strut{}Internal magnetic field H (kA/m)}}%
    }%
    \gplgaddtomacro\gplfronttext{%
    }%
    \gplbacktext
    \put(0,0){\includegraphics{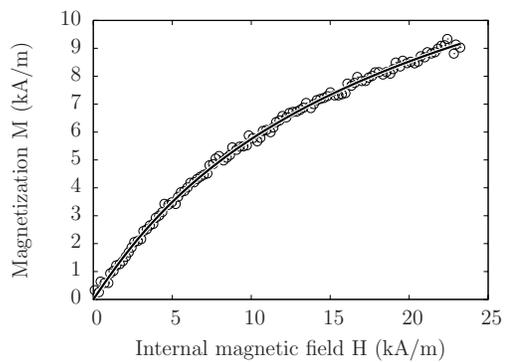}}%
    \gplfronttext
  \end{picture}%
\endgroup
}
\caption{\label{fig:ballmag} Magnetization curve of the ball.}
\end{figure}

\subsection{Magnetization}
Next, we measure the magnetization of the sphere for various fields, utilizing a fluxmetric
magnetometer (Lakeshore, Model 480). The resulting magnetization curve is plotted in
Fig.~\ref{fig:ballmag}. The solid line displays an approximation with the model presented in Ref.~\cite{ivanov2001}.
The sample is superparamagnetic with an initial
susceptibility $\chi_0=0.81$.

\begin{figure}
\resizebox{0.9\columnwidth}{!}{\grfsize
\begingroup 
  \gdef\gplbacktext{}%
  \gdef\gplfronttext{}%

  \ifGPblacktext
    \def\colorrgb#1{}%
    \def\colorgray#1{}%
  \else
    \ifGPcolor
      \def\colorrgb#1{\color[rgb]{#1}}%
      \def\colorgray#1{\color[gray]{#1}}%
      \expandafter\def\csname LTw\endcsname{\color{white}}%
      \expandafter\def\csname LTb\endcsname{\color{black}}%
      \expandafter\def\csname LTa\endcsname{\color{black}}%
      \expandafter\def\csname LT0\endcsname{\color[rgb]{1,0,0}}%
      \expandafter\def\csname LT1\endcsname{\color[rgb]{0,1,0}}%
      \expandafter\def\csname LT2\endcsname{\color[rgb]{0,0,1}}%
      \expandafter\def\csname LT3\endcsname{\color[rgb]{1,0,1}}%
      \expandafter\def\csname LT4\endcsname{\color[rgb]{0,1,1}}%
      \expandafter\def\csname LT5\endcsname{\color[rgb]{1,1,0}}%
      \expandafter\def\csname LT6\endcsname{\color[rgb]{0,0,0}}%
      \expandafter\def\csname LT7\endcsname{\color[rgb]{1,0.3,0}}%
      \expandafter\def\csname LT8\endcsname{\color[rgb]{0.5,0.5,0.5}}%
    \else
      \def\colorrgb#1{\color{black}}%
      \def\colorgray#1{\color[gray]{#1}}%
      \expandafter\def\csname LTw\endcsname{\color{white}}%
      \expandafter\def\csname LTb\endcsname{\color{black}}%
      \expandafter\def\csname LTa\endcsname{\color{black}}%
      \expandafter\def\csname LT0\endcsname{\color{black}}%
      \expandafter\def\csname LT1\endcsname{\color{black}}%
      \expandafter\def\csname LT2\endcsname{\color{black}}%
      \expandafter\def\csname LT3\endcsname{\color{black}}%
      \expandafter\def\csname LT4\endcsname{\color{black}}%
      \expandafter\def\csname LT5\endcsname{\color{black}}%
      \expandafter\def\csname LT6\endcsname{\color{black}}%
      \expandafter\def\csname LT7\endcsname{\color{black}}%
      \expandafter\def\csname LT8\endcsname{\color{black}}%
    \fi
  \fi
  \setlength{\unitlength}{0.0500bp}%
  \begin{picture}(7200.00,5040.00)%
    \gplgaddtomacro\gplbacktext{%
      \csname LTb\endcsname%
      \put(1428,840){\makebox(0,0)[r]{\strut{} 0}}%
      \put(1428,1806){\makebox(0,0)[r]{\strut{} 400}}%
      \put(1428,2772){\makebox(0,0)[r]{\strut{} 800}}%
      \put(1428,3738){\makebox(0,0)[r]{\strut{} 1200}}%
      \put(1428,4704){\makebox(0,0)[r]{\strut{} 1600}}%
      \put(1596,560){\makebox(0,0){\strut{} 0.001}}%
      \put(2451,560){\makebox(0,0){\strut{} 0.01}}%
      \put(3305,560){\makebox(0,0){\strut{} 0.1}}%
      \put(4160,560){\makebox(0,0){\strut{} 1}}%
      \put(5015,560){\makebox(0,0){\strut{} 10}}%
      \put(5869,560){\makebox(0,0){\strut{} 100}}%
      \put(6724,560){\makebox(0,0){\strut{} 1000}}%
      \put(280,2772){\rotatebox{90}{\makebox(0,0){\strut{}Shear modulus $G$ (Pa)}}}%
      \put(4160,140){\makebox(0,0){\strut{}Elapsed time (s)}}%
    }%
    \gplgaddtomacro\gplfronttext{%
    }%
    \gplbacktext
    \put(0,0){\includegraphics{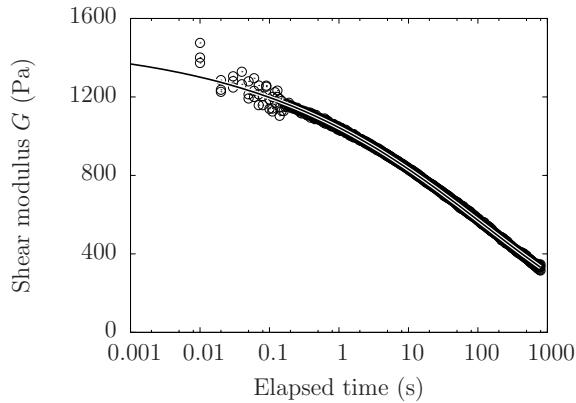}}%
    \gplfronttext
  \end{picture}%
\endgroup
}
\caption{The force response to a jump in the deformation. The solid line
displays a fit to Eq.~(\ref{eq:kww}) with the parameters $G_\text{rheo}=1480\,\mathrm{Pa}$,
$t_0=127\,\,\mathrm{s}$ and $\beta=0.218$.}
\label{fig:creep}
\end{figure}

\subsection{Rheology}
For the characterization of the elastic properties we measure the shear modulus $G$.
with a rheometer (MCR\,301, Anton Paar) in cone and plate geometry. The cone has a
diameter of $50\,\mathrm{mm}$ and a base angle of $1$\textdegree. We perform a
stress relaxation experiment on the sample: from the equilibrium position we
shear the sample by a deformation of $\gamma=1\,\%$ and measure the stress $\tau$ as a function of time.
Figure~\ref{fig:creep} displays the results. The restoring force decays by $50\,\%$ during one
second, which means that the shear modulus $G=\tau/\gamma$ cannot be treated as constant.
This decay can well be approximated by a stretched exponential (the solid line in
Fig.~\ref{fig:creep})
\begin{equation}
G(t) = G_\text{rheo} \exp\left( - (\nicefrac{t}{t_0})^\beta\right).
\label{eq:kww}
\end{equation}
The material therefore softens under load. To account for that time dependence, the magnetic
experiment cannot be performed in a static manner.

\begin{figure}
\centering
\resizebox{0.9\columnwidth}{!}{\grfsize
\begingroup 
  \gdef\gplbacktext{}%
  \gdef\gplfronttext{}%

   \ifGPblacktext
    \def\colorrgb#1{}%
    \def\colorgray#1{}%
  \else
    \ifGPcolor
      \def\colorrgb#1{\color[rgb]{#1}}%
      \def\colorgray#1{\color[gray]{#1}}%
      \expandafter\def\csname LTw\endcsname{\color{white}}%
      \expandafter\def\csname LTb\endcsname{\color{black}}%
      \expandafter\def\csname LTa\endcsname{\color{black}}%
      \expandafter\def\csname LT0\endcsname{\color[rgb]{1,0,0}}%
      \expandafter\def\csname LT1\endcsname{\color[rgb]{0,1,0}}%
      \expandafter\def\csname LT2\endcsname{\color[rgb]{0,0,1}}%
      \expandafter\def\csname LT3\endcsname{\color[rgb]{1,0,1}}%
      \expandafter\def\csname LT4\endcsname{\color[rgb]{0,1,1}}%
      \expandafter\def\csname LT5\endcsname{\color[rgb]{1,1,0}}%
      \expandafter\def\csname LT6\endcsname{\color[rgb]{0,0,0}}%
      \expandafter\def\csname LT7\endcsname{\color[rgb]{1,0.3,0}}%
      \expandafter\def\csname LT8\endcsname{\color[rgb]{0.5,0.5,0.5}}%
    \else
      \def\colorrgb#1{\color{black}}%
      \def\colorgray#1{\color[gray]{#1}}%
      \expandafter\def\csname LTw\endcsname{\color{white}}%
      \expandafter\def\csname LTb\endcsname{\color{black}}%
      \expandafter\def\csname LTa\endcsname{\color{black}}%
      \expandafter\def\csname LT0\endcsname{\color{black}}%
      \expandafter\def\csname LT1\endcsname{\color{black}}%
      \expandafter\def\csname LT2\endcsname{\color{black}}%
      \expandafter\def\csname LT3\endcsname{\color{black}}%
      \expandafter\def\csname LT4\endcsname{\color{black}}%
      \expandafter\def\csname LT5\endcsname{\color{black}}%
      \expandafter\def\csname LT6\endcsname{\color{black}}%
      \expandafter\def\csname LT7\endcsname{\color{black}}%
      \expandafter\def\csname LT8\endcsname{\color{black}}%
    \fi
  \fi
  \setlength{\unitlength}{0.0500bp}%
  \begin{picture}(7200.00,5040.00)%
    \gplgaddtomacro\gplbacktext{%
      \csname LTb\endcsname%
      \put(1260,840){\makebox(0,0)[r]{\strut{} 0}}%
      \put(1260,1613){\makebox(0,0)[r]{\strut{} 0.5}}%
      \put(1260,2386){\makebox(0,0)[r]{\strut{} 1}}%
      \put(1260,3158){\makebox(0,0)[r]{\strut{} 1.5}}%
      \put(1260,3931){\makebox(0,0)[r]{\strut{} 2}}%
      \put(1260,4704){\makebox(0,0)[r]{\strut{} 2.5}}%
      \put(1428,560){\makebox(0,0){\strut{} 0.001}}%
      \put(2835,560){\makebox(0,0){\strut{} 0.01}}%
      \put(4242,560){\makebox(0,0){\strut{} 0.1}}%
      \put(5649,560){\makebox(0,0){\strut{} 1}}%
      \put(280,2772){\rotatebox{90}{\makebox(0,0){\strut{}Elongation $\varepsilon$ (\%)}}}%
      \put(7199,2772){\rotatebox{90}{\makebox(0,0){\strut{}}}}%
      \put(4174,140){\makebox(0,0){\strut{}Time (s)}}%
      \put(4174,4564){\makebox(0,0){\strut{}}}%
      \put(4174,4563){\makebox(0,0){\strut{}}}%
      \put(420,140){\makebox(0,0)[l]{\strut{}}}%
    }%
    \gplgaddtomacro\gplfronttext{%
    }%
    \gplbacktext
    \put(0,0){\includegraphics{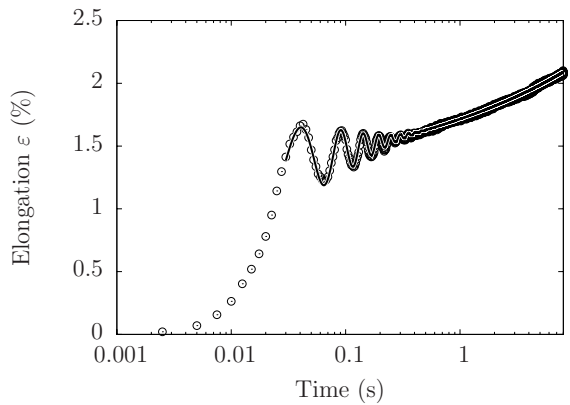}}%
    \gplfronttext
  \end{picture}%
\endgroup
}
\caption{Elastic response of the ball, when a magnetic field is suddenly
applied. The solid line represents a fit to Eq.~(\ref{eq:harmosc}).}
\label{fig:pulsresponse}
\end{figure}

\section{Results and discussion}
We performed time-resolved measurements of the relative elongation $\varepsilon$ of the ball after applying
a magnetic induction $B$ in a jumplike fashion for ten different values of $B$. 
Figure~\ref{fig:pulsresponse} shows $\varepsilon(t)$ for the case $B=36\,\mathrm{mT}$. The
elongation $\varepsilon=(d-d_0)/d_0$
measures the scaled difference of the diameters in the direction of the field~($d$) and without a
field~($d_0$). Due to the sudden increase of the magnetic field and the inertia of the ball,
the latter performs uniaxial damped vibrations with a frequency
$f=22.4\,\mathrm{Hz}$. When the oscillations cease, the elongation continues to
grow due to the softening under load. Because the timescale of the softening is much 
larger than that of the oscillation, the experiment may be approximated by a
harmonic oscillator pulled by a constant force, where the spring constant
relaxes according to Eq.~(\ref{eq:kww})
\begin{equation}
 \ddot \varepsilon + \delta \dot \varepsilon + \omega_0^2  \exp\left( -
    (\nicefrac{t}{t_0})^\beta\right)\,\varepsilon
        = F.
\label{eq:harmosc}
\end{equation}
Here $\delta$ is the damping constant, $F$ the pulling force (i.e. related to
the magnetic induction) and $\omega_0$ the natural frequency. A fit with that
equation is displayed in Fig.~\ref{fig:pulsresponse} as the solid line.
The ratio $F/\omega_0^2$
corresponds to the relative elongation in the equilibrium state if the spring constant would not change.
Therefore, this ratio compares to the elongation predicted by the static theories.

\begin{figure}
\centering
\resizebox{\columnwidth}{!}{\grfsize\fontsize{13.5pt}{13.5pt}\selectfont
\begingroup
  \gdef\gplbacktext{}%
  \gdef\gplfronttext{}%
  \ifGPblacktext
    \def\colorrgb#1{}%
    \def\colorgray#1{}%
  \else
    \ifGPcolor
      \def\colorrgb#1{\color[rgb]{#1}}%
      \def\colorgray#1{\color[gray]{#1}}%
      \expandafter\def\csname LTw\endcsname{\color{white}}%
      \expandafter\def\csname LTb\endcsname{\color{black}}%
      \expandafter\def\csname LTa\endcsname{\color{black}}%
      \expandafter\def\csname LT0\endcsname{\color[rgb]{1,0,0}}%
      \expandafter\def\csname LT1\endcsname{\color[rgb]{0,1,0}}%
      \expandafter\def\csname LT2\endcsname{\color[rgb]{0,0,1}}%
      \expandafter\def\csname LT3\endcsname{\color[rgb]{1,0,1}}%
      \expandafter\def\csname LT4\endcsname{\color[rgb]{0,1,1}}%
      \expandafter\def\csname LT5\endcsname{\color[rgb]{1,1,0}}%
      \expandafter\def\csname LT6\endcsname{\color[rgb]{0,0,0}}%
      \expandafter\def\csname LT7\endcsname{\color[rgb]{1,0.3,0}}%
      \expandafter\def\csname LT8\endcsname{\color[rgb]{0.5,0.5,0.5}}%
    \else
      \def\colorrgb#1{\color{black}}%
      \def\colorgray#1{\color[gray]{#1}}%
      \expandafter\def\csname LTw\endcsname{\color{white}}%
      \expandafter\def\csname LTb\endcsname{\color{black}}%
      \expandafter\def\csname LTa\endcsname{\color{black}}%
      \expandafter\def\csname LT0\endcsname{\color{black}}%
      \expandafter\def\csname LT1\endcsname{\color{black}}%
      \expandafter\def\csname LT2\endcsname{\color{black}}%
      \expandafter\def\csname LT3\endcsname{\color{black}}%
      \expandafter\def\csname LT4\endcsname{\color{black}}%
      \expandafter\def\csname LT5\endcsname{\color{black}}%
      \expandafter\def\csname LT6\endcsname{\color{black}}%
      \expandafter\def\csname LT7\endcsname{\color{black}}%
      \expandafter\def\csname LT8\endcsname{\color{black}}%
    \fi
  \fi
  \setlength{\unitlength}{0.0500bp}%
  \begin{picture}(7200.00,7200.00)%
    \gplgaddtomacro\gplbacktext{%
      \csname LTb\endcsname%
      \put(1512,3376){\makebox(0,0)[r]{\strut{}-1}}%
      \put(1512,4074){\makebox(0,0)[r]{\strut{}-0.5}}%
      \put(1512,4771){\makebox(0,0)[r]{\strut{} 0}}%
      \put(1512,5469){\makebox(0,0)[r]{\strut{} 0.5}}%
      \put(1512,6166){\makebox(0,0)[r]{\strut{} 1}}%
      \put(1512,6864){\makebox(0,0)[r]{\strut{} 1.5}}%
      \put(1960,3096){\makebox(0,0){\strut{}}}%
      \put(2521,3096){\makebox(0,0){\strut{}}}%
      \put(3081,3096){\makebox(0,0){\strut{}}}%
      \put(3642,3096){\makebox(0,0){\strut{}}}%
      \put(4202,3096){\makebox(0,0){\strut{}}}%
      \put(4762,3096){\makebox(0,0){\strut{}}}%
      \put(5323,3096){\makebox(0,0){\strut{}}}%
      \put(5883,3096){\makebox(0,0){\strut{}}}%
      \put(6444,3096){\makebox(0,0){\strut{}}}%
      \csname LTb\endcsname%
      \put(532,5120){\rotatebox{90}{\makebox(0,0){\strut{}Elongation $\varepsilon (\%)$}}}%
      \put(0,3024){\makebox(0,0)[l]{\strut{}(a)}}%
    }%
    \gplgaddtomacro\gplfronttext{%
    }%
    \gplgaddtomacro\gplbacktext{%
      \csname LTb\endcsname%
      \put(1512,1500){\makebox(0,0)[r]{\strut{} 20}}%
      \put(1512,2073){\makebox(0,0)[r]{\strut{} 21}}%
      \put(1512,2647){\makebox(0,0)[r]{\strut{} 22}}%
      \put(1960,704){\makebox(0,0){\strut{}10}}%
      \put(2521,704){\makebox(0,0){\strut{}20}}%
      \put(3081,704){\makebox(0,0){\strut{}30}}%
      \put(3642,704){\makebox(0,0){\strut{}40}}%
      \put(4202,704){\makebox(0,0){\strut{}50}}%
      \put(4762,704){\makebox(0,0){\strut{}60}}%
      \put(5323,704){\makebox(0,0){\strut{}70}}%
      \put(5883,704){\makebox(0,0){\strut{}80}}%
      \put(6444,704){\makebox(0,0){\strut{}90}}%
      \put(700,2016){\rotatebox{90}{\makebox(0,0){\strut{}Frequency $f$\,(Hz)}}}%
      \put(4202,284){\makebox(0,0){\strut{}Magnetization $M^2\, (\mathrm{kA/m})$}}%
      \put(0,0){\makebox(0,0)[l]{\strut{}(b)}}%
    }%
    \gplgaddtomacro\gplfronttext{%
    }%
    \gplbacktext
    \put(0,0){\includegraphics{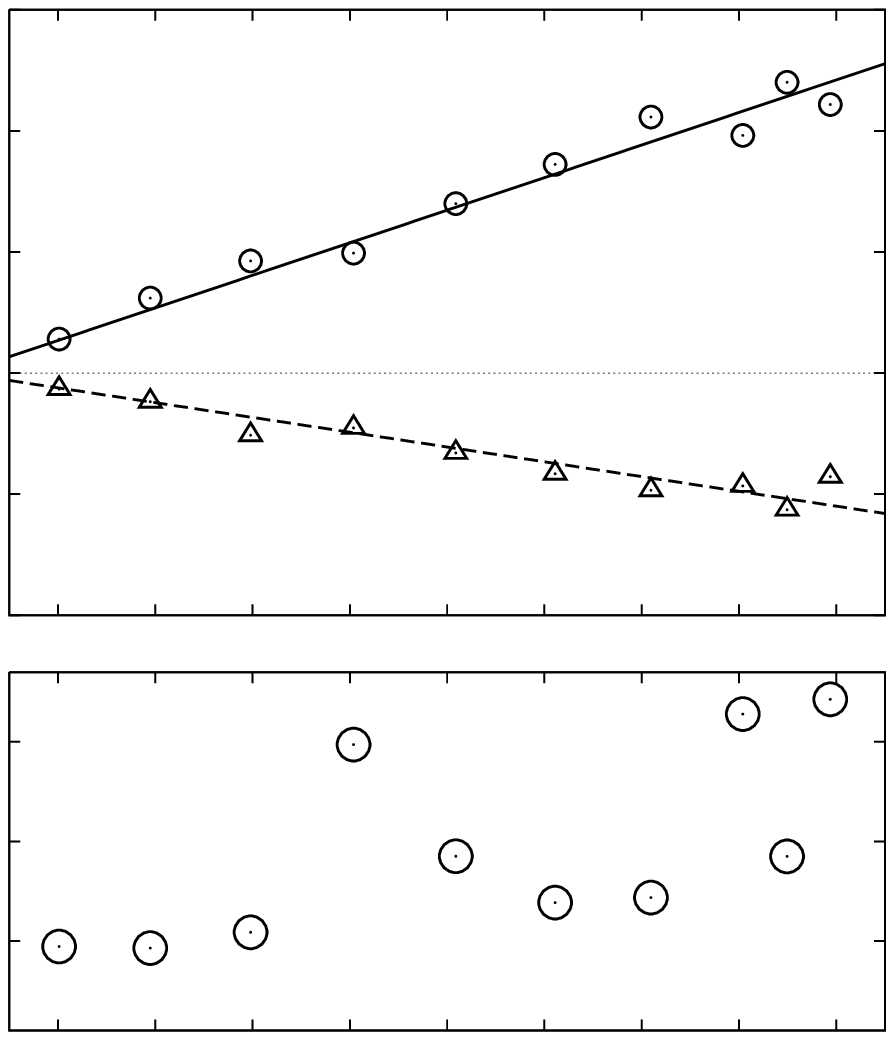}}%
    \gplfronttext
  \end{picture}%
\endgroup
}
\caption{a) The static elongation as a function of $M^2$. The circles~(triangles) denote the
elongation~(contraction) parallel~(perpendicular) to the applied field, respectively. The solid~(dashed) line is the best linear fit.
b) The frequencies  of the initial vibrations obtained from Eq.~(\ref{eq:harmosc}).}
\label{fig:elfreq}
\end{figure}

The dependence of the elongation on the magnetization is shown in Fig.~\ref{fig:elfreq}\,a. For ten different
values of the magnetic induction we have recorded and evaluated the elongation $\varepsilon$ of the
ball. Figure~\ref{fig:elfreq}\,a presents the outcome for the elongation parallel~($\varepsilon_z$) and
perpendicular~($\varepsilon_x$) to the magnetic field. The data have been plotted versus $M^2$. In
agreement with Eq.~(\ref{eq:relelong}) we find a linear relationship $\varepsilon_i=c_i M^2$ with  
$c_z=13.4\cdot10^{-5}(\mathrm{\frac{m}{kA}})^2$ and $c_x=-6.1\cdot10^{-5}(\mathrm{\frac{m}{kA}})^2$.
Figure~\ref{fig:elfreq}\,b shows the oscillation frequency $f=\omega_0/(2\pi)$ versus $M^2$.

Next we compare the ratio of the slopes $c_z/c_x=\kappa_z/\kappa_x$ with the theoretical predictions. 
For a deformation restricted to an ellipsoidal shape and the assumption of a uniform strain field,
the expressions given in \cite{landau1960_8,raikher2005} can be rewritten in terms of
Poisson's ratio $\sigma$
\begin{subequations}
\begin{align}
\kappa_z =& \frac{3-2\sigma}{20\sigma+20}\\
\kappa_x =& \frac{1-4\sigma}{20\sigma+20}.
\label{eq:landaukappa}
\end{align}
\end{subequations}
For an incompressible gel ($\sigma=1/2$) this leads to $\kappa_z=\frac{1}{15}$. 
For arbitrary $\sigma$ one obtains for the ratio
\begin{equation}
\kappa_x/\kappa_z = \frac{1 - 4\sigma}{3 - 2\sigma}. 
\label{eq:landaurat}
\end{equation}
We exploit Eq.~(\ref{eq:landaurat}) to
determine $\sigma$, and by substituting $\sigma$ in Eqs.~(\ref{eq:landaukappa},\ref{eq:relelong})
one finally arrives at $G$. This yields
\begin{equation}
\sigma=0.48\pm0.01 \quad G=0.65\,\mathrm{kPa}.
\end{equation}

For the more general case of a non-uniform strain field and a shape not restricted to an
ellipsoid~\cite{raikher2005}, one obtains
\begin{subequations}
\begin{align}
\kappa_z =& -\frac{6\sigma^2+\sigma-7}{20\sigma^2+48\sigma+28}\\
\kappa_x =& -\frac{\sigma^2+2\sigma}{10\sigma^2+24\sigma+14} 
\label{eq:kappa}
\end{align}
\end{subequations}
which yields
\begin{equation}
\kappa_x/\kappa_z = \frac{2\sigma^2+4\sigma}{6\sigma^2+\sigma-7}
\end{equation}
and finally gives 
\begin{equation}
\sigma=0.47\pm0.01 \quad G=0.87\,\mathrm{kPa}.
\end{equation}

\begin{figure}
\centering
\resizebox{0.85\columnwidth}{!}{\grfsize
\begingroup
  \gdef\gplbacktext{}%
  \gdef\gplfronttext{}%
  \ifGPblacktext
    \def\colorrgb#1{}%
    \def\colorgray#1{}%
  \else
    \ifGPcolor
      \def\colorrgb#1{\color[rgb]{#1}}%
      \def\colorgray#1{\color[gray]{#1}}%
      \expandafter\def\csname LTw\endcsname{\color{white}}%
      \expandafter\def\csname LTb\endcsname{\color{black}}%
      \expandafter\def\csname LTa\endcsname{\color{black}}%
      \expandafter\def\csname LT0\endcsname{\color[rgb]{1,0,0}}%
      \expandafter\def\csname LT1\endcsname{\color[rgb]{0,1,0}}%
      \expandafter\def\csname LT2\endcsname{\color[rgb]{0,0,1}}%
      \expandafter\def\csname LT3\endcsname{\color[rgb]{1,0,1}}%
      \expandafter\def\csname LT4\endcsname{\color[rgb]{0,1,1}}%
      \expandafter\def\csname LT5\endcsname{\color[rgb]{1,1,0}}%
      \expandafter\def\csname LT6\endcsname{\color[rgb]{0,0,0}}%
      \expandafter\def\csname LT7\endcsname{\color[rgb]{1,0.3,0}}%
      \expandafter\def\csname LT8\endcsname{\color[rgb]{0.5,0.5,0.5}}%
    \else
      \def\colorrgb#1{\color{black}}%
      \def\colorgray#1{\color[gray]{#1}}%
      \expandafter\def\csname LTw\endcsname{\color{white}}%
      \expandafter\def\csname LTb\endcsname{\color{black}}%
      \expandafter\def\csname LTa\endcsname{\color{black}}%
      \expandafter\def\csname LT0\endcsname{\color{black}}%
      \expandafter\def\csname LT1\endcsname{\color{black}}%
      \expandafter\def\csname LT2\endcsname{\color{black}}%
      \expandafter\def\csname LT3\endcsname{\color{black}}%
      \expandafter\def\csname LT4\endcsname{\color{black}}%
      \expandafter\def\csname LT5\endcsname{\color{black}}%
      \expandafter\def\csname LT6\endcsname{\color{black}}%
      \expandafter\def\csname LT7\endcsname{\color{black}}%
      \expandafter\def\csname LT8\endcsname{\color{black}}%
    \fi
  \fi
  \setlength{\unitlength}{0.0500bp}%
  \begin{picture}(7200.00,5040.00)%
    \gplgaddtomacro\gplbacktext{%
      \csname LTb\endcsname%
      \put(1428,840){\makebox(0,0)[r]{\strut{} 0}}%
      \put(1428,1323){\makebox(0,0)[r]{\strut{} 200}}%
      \put(1428,1806){\makebox(0,0)[r]{\strut{} 400}}%
      \put(1428,2289){\makebox(0,0)[r]{\strut{} 600}}%
      \put(1428,2772){\makebox(0,0)[r]{\strut{} 800}}%
      \put(1428,3255){\makebox(0,0)[r]{\strut{} 1000}}%
      \put(1428,3738){\makebox(0,0)[r]{\strut{} 1200}}%
      \put(1428,4221){\makebox(0,0)[r]{\strut{} 1400}}%
      \put(1428,4704){\makebox(0,0)[r]{\strut{} 1600}}%
      \put(1596,560){\makebox(0,0){\strut{} 5}}%
      \put(2329,560){\makebox(0,0){\strut{} 10}}%
      \put(3061,560){\makebox(0,0){\strut{} 15}}%
      \put(3794,560){\makebox(0,0){\strut{} 20}}%
      \put(4526,560){\makebox(0,0){\strut{} 25}}%
      \put(5259,560){\makebox(0,0){\strut{} 30}}%
      \put(5991,560){\makebox(0,0){\strut{} 35}}%
      \put(6724,560){\makebox(0,0){\strut{} 40}}%
      \csname LTb\endcsname%
      \put(280,2772){\rotatebox{90}{\makebox(0,0){\strut{}Shear modulus $G$\,(Pa)}}}%
      \put(4160,140){\makebox(0,0){\strut{}Magnetic induction $B\,(\mathrm{mT})$}}%
    }%
    \gplgaddtomacro\gplfronttext{%
    }%
    \gplbacktext
    \put(0,0){\includegraphics{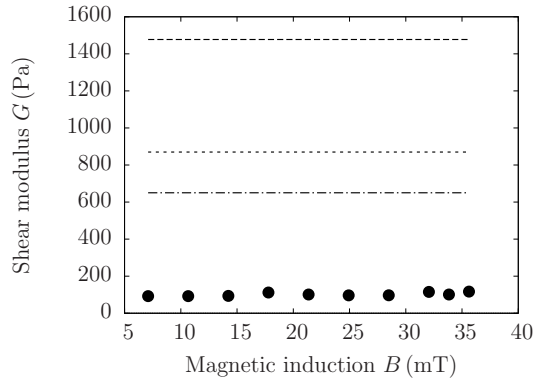}}%
    \gplfronttext
  \end{picture}%
\endgroup
}
\caption{The shear modulus measured by different methods: Commercial rheometer (dashed),
elongation theory\cite{raikher2005} (short dashed), elongation theory\cite{landau1960_8}(dash-dotted),
vibrations (solid circles).}
\label{fig:gcompare}
\end{figure}

Both approaches yield a Poisson's ratio $\sigma$ very close to the limit of incompressibility
$\sigma=\nicefrac{1}{2}$, which is characteristic for rubberlike materials~\cite{rinde1970}.
As reported before~\cite{raikher2005}, the values derived for the shear modulus $G$ differ by
$\approx 30\,\%$.  
But the value obtained from the rheometer,
$G_\text{rheo}=1.48\,\mathrm{kPa}$, exceeds the old and new predictions by a factor of $2.3$ and
$1.7$, respectively. So none of them is corroborated by the experiment.

The frequency of the vibrations after the sudden increase of the magnetic field offers another
possibility to measure the shear modulus. For an incompressible elastic sphere that performs spheroidal
vibrations, where the sphere gets alternately deformed into a prolate and oblate ellipsoid of
revolution, the frequency is given~\cite{love1944} by
\begin{equation}
f = 0.848 \sqrt{\frac{G}{4\rho r^2}}.
\end{equation}
Since the radius of the sphere $r$ and the density $\rho$ are known, we can compute the shear
modulus from the measured vibration frequency $f$. Figure~\ref{fig:gcompare} shows a comparison of the
value of the shear modulus obtained from $f$ with the values from the
elongation and the mechanical measurement. The average shear modulus determined by this method is 
$G_\text{vib}=0.1\,\mathrm{kPa}$, which differs by a factor of $15$ from $G_\text{rheo}$. 
This large deviation may arise, because the model for the vibrations does not take into account the
surrounding water. This needs to oscillate together with the sphere, leading to an increased
effective mass of the oscillator, and thus a reduced frequency. 

\begin{figure}
\centering
\resizebox{0.9\columnwidth}{!}{\grfsize
\begingroup
  \gdef\gplbacktext{}%
  \gdef\gplfronttext{}%
  \ifGPblacktext
    \def\colorrgb#1{}%
    \def\colorgray#1{}%
  \else
    \ifGPcolor
      \def\colorrgb#1{\color[rgb]{#1}}%
      \def\colorgray#1{\color[gray]{#1}}%
      \expandafter\def\csname LTw\endcsname{\color{white}}%
      \expandafter\def\csname LTb\endcsname{\color{black}}%
      \expandafter\def\csname LTa\endcsname{\color{black}}%
      \expandafter\def\csname LT0\endcsname{\color[rgb]{1,0,0}}%
      \expandafter\def\csname LT1\endcsname{\color[rgb]{0,1,0}}%
      \expandafter\def\csname LT2\endcsname{\color[rgb]{0,0,1}}%
      \expandafter\def\csname LT3\endcsname{\color[rgb]{1,0,1}}%
      \expandafter\def\csname LT4\endcsname{\color[rgb]{0,1,1}}%
      \expandafter\def\csname LT5\endcsname{\color[rgb]{1,1,0}}%
      \expandafter\def\csname LT6\endcsname{\color[rgb]{0,0,0}}%
      \expandafter\def\csname LT7\endcsname{\color[rgb]{1,0.3,0}}%
      \expandafter\def\csname LT8\endcsname{\color[rgb]{0.5,0.5,0.5}}%
    \else
      \def\colorrgb#1{\color{black}}%
      \def\colorgray#1{\color[gray]{#1}}%
      \expandafter\def\csname LTw\endcsname{\color{white}}%
      \expandafter\def\csname LTb\endcsname{\color{black}}%
      \expandafter\def\csname LTa\endcsname{\color{black}}%
      \expandafter\def\csname LT0\endcsname{\color{black}}%
      \expandafter\def\csname LT1\endcsname{\color{black}}%
      \expandafter\def\csname LT2\endcsname{\color{black}}%
      \expandafter\def\csname LT3\endcsname{\color{black}}%
      \expandafter\def\csname LT4\endcsname{\color{black}}%
      \expandafter\def\csname LT5\endcsname{\color{black}}%
      \expandafter\def\csname LT6\endcsname{\color{black}}%
      \expandafter\def\csname LT7\endcsname{\color{black}}%
      \expandafter\def\csname LT8\endcsname{\color{black}}%
    \fi
  \fi
  \setlength{\unitlength}{0.0500bp}%
  \begin{picture}(7200.00,5040.00)%
    \gplgaddtomacro\gplbacktext{%
      \csname LTb\endcsname%
      \put(1260,840){\makebox(0,0)[r]{\strut{} 0}}%
      \put(1260,1458){\makebox(0,0)[r]{\strut{} 0.2}}%
      \put(1260,2076){\makebox(0,0)[r]{\strut{} 0.4}}%
      \put(1260,2695){\makebox(0,0)[r]{\strut{} 0.6}}%
      \put(1260,3313){\makebox(0,0)[r]{\strut{} 0.8}}%
      \put(1260,3931){\makebox(0,0)[r]{\strut{} 1}}%
      \put(1260,4549){\makebox(0,0)[r]{\strut{} 1.2}}%
      \put(1428,560){\makebox(0,0){\strut{}-1000}}%
      \put(2090,560){\makebox(0,0){\strut{} 0}}%
      \put(2752,560){\makebox(0,0){\strut{} 1000}}%
      \put(3414,560){\makebox(0,0){\strut{} 2000}}%
      \put(4076,560){\makebox(0,0){\strut{} 3000}}%
      \put(4738,560){\makebox(0,0){\strut{} 4000}}%
      \put(5400,560){\makebox(0,0){\strut{} 5000}}%
      \put(6062,560){\makebox(0,0){\strut{} 6000}}%
      \put(6724,560){\makebox(0,0){\strut{} 7000}}%
      \csname LTb\endcsname%
      \put(280,2772){\rotatebox{90}{\makebox(0,0){\strut{}Deformation $\gamma$}}}%
      \put(4076,140){\makebox(0,0){\strut{}Elapsed time (s)}}%
    }%
    \gplgaddtomacro\gplfronttext{%
    }%
    \gplbacktext
    \put(0,0){\includegraphics{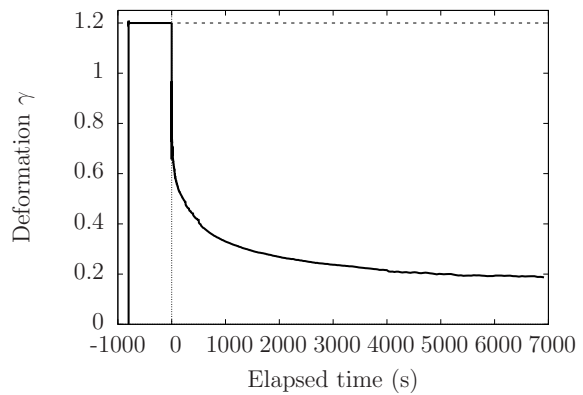}}%
    \gplfronttext
  \end{picture}%
\endgroup
}
\caption{The shear deformation as a function of time, after applying a constant deformation of
$1.2\,\%$ for $800\,\mathrm{s}$.}
\label{fig:stressrelease}
\end{figure}

Now we reconsider the deviation between the shear modulus $G$ as determined from the elongations and
$G_\text{rheo}$ determined via rheology. They differ by factor of about two. 
This difference cannot be explained solely by the inaccuracy of
the measurement of $G_\text{rheo}$, which is typically around $10\,\%$ for a commercial rheometer.
Rather, the deviation is likely to stem from the fact that the idealistic
models~\cite{landau1960_8,raikher2005} consider 
only Hookean elasticity, which is insufficient for our material. 
In fact, the deformation of a gel put under load
can be made up of three contributions, namely instantaneous elastic deformation, retarded anelastic
deformation and viscous flow~\cite{strobl1997}. Our material clearly shows anelasticity, as
demonstrated in Fig.~\ref{fig:creep}. 
Moreover, our sample shows the phenomenon of viscous flow, as illustrated in
Fig.~\ref{fig:stressrelease}.  In this measurement we first apply a constant strain for
$800\,\mathrm{s}$, and then record the relaxation of the strain for zero stress. The remaining
deformation at $t=7000\,\mathrm{s}$ is about $15\,\%$ of the strain initially applied.  This value
can be regarded as an upper bound for the plastic contribution to the deformation (viscous flow).
While the existence of anelasticity and viscous flow gives no straight-forward explanation of the
$70\,\%$ deviation between experiment and theory, it is obvious that the full viscoelastic behaviour
of the gel should be included in the computations from the very beginning. 

\section{Conclusion}
We have measured the deformation of a ferrogel sphere in response to a uniform magnetic field by
direct optical means. We compare the results for the first time with the models in Refs.~\cite{landau1960_8,raikher2005}. 
From the ratio of the elongation parallel and perpendicular to the field, we calculate
Poisson's ratio, which is close to the value $\sigma=0.5$ expected for an incompressible material. 
More importantly, the absolute value of the elongation is $70\,\%$ larger than the one calculated from the
models.  This is presumable caused by the neglection of the anelasticity and viscous flow of the
ferrogel.
So further theoretical investigation, including the full viscoelastic properties of the gel, is
needed to explain the experiment quantitatively. With such a model one would also be able to compute the
dynamic response of the gel under a sudden change of the external magnetic field.  This is not
only of fundamental interest, but also important for possible technical applications of these smart
materials. 

\section{Acknowledgments}
The authors thank Helmut R.~Brand and Yurij L.~Raikher for fruitful discussions. Financial support
by the Deutsche Forschungsgemeinschaft via  FOR~608 is gratefully acknowledged. A.T.
is grateful for an INTAS Young Scientists Fellowship (05-109-4521).
We thank Total (Finavestan A~50\,B) and Kraton Polymers (Kraton G-1650) for providing samples.

\bibliographystyle{apsrev} 
\bibliography{alles}
\end{document}